\documentstyle[12pt,aps,prc,epsfig,preprint]{revtex}
\tightenlines
\begin{document}
\begin{titlepage}
\title{\vspace*{10mm}\bf
\Large $K^-$ - nucleus relativistic mean field potentials
consistent with kaonic atoms}
\vspace{6pt}

\author{  E.~Friedman$^a$, A.~Gal$^a$, J.~Mare\v{s}$^b$ ,
A.~Ciepl\'{y}$^{a,b}$\\
$^a${\it Racah Institute of Physics, The Hebrew University, Jerusalem 91904,
Israel\\}
$^b${\it Nuclear Physics Institute, 25068 \v{R}e\v{z}, Czech Republic}}

\vspace{4pt}
\maketitle

\begin{abstract}
$K^-$ atomic data are used to test several models of the $K^-$ nucleus
interaction. The t($\rho$)$\rho$ optical potential, due to coupled channel
models incorporating the $\Lambda$(1405) dynamics, fails to reproduce these data.
A standard relativistic mean field (RMF) potential, disregarding the
$\Lambda$(1405) dynamics at low densities, also fails. The only successful
model is a hybrid of a theoretically motivated RMF approach in the nuclear
interior and a completely phenomenological density dependent potential,
which respects the low density theorem in the nuclear surface region.
This best-fit $K^-$ optical potential is found to be strongly attractive,
with a depth of 180$\pm$20~MeV at the nuclear interior, in agreement with
previous phenomenological analyses.
\newline
$PACS$: 24.10.Ht; 36.10.Gv
\end{abstract}\vspace{1cm}
\centerline{\today}
\end{titlepage}

\section{Introduction}
\label{sec:int}

The interaction and properties of $K^-$ mesons in nuclear matter have
attracted considerable attention in the last few years, as may be gathered
from the recent reviews in the Dover memorial volume of Nuclear Physics A
\cite{ww97,smb97,llb97} and the references cited therein. Among the topics
discussed was the question of kaon condensation \cite{kn86} in dense
nuclear matter and the possibility of the onset of such condensation
in neutron stars \cite{blrt94,tpl94,sm96}. Formation of the condensate
would soften the equation of state of baryonic matter and consequently
would reduce the upper limit for the mass of neutron stars.
Such scenarios depend  on the depth of the attractive
$K^-$ optical potential at normal nuclear densities where
experimental information is available.

Theoretical considerations based on One Boson Exchange (OBE) models,
coupled channel chiral perturbation theory approach, or Relativistic
Mean Field (RMF) models, yield for the attractive real part of the
$K^-$ nucleus potential at threshold depths in the range
between 70 and 120~MeV \cite{ww97,smb97,llb97}.
The reason for such a  spread of the predicted values lies in
the diversity of treating the $\Lambda$(1405) hyperon, which is
considered to be an unstable $\bar KN$ bound state just below the
$K^-p$ threshold. The purpose of the present work is to test some
of these  $K^-$ nucleus interaction models by confronting
these models with kaonic atom data. This is particularly topical,
since recent fits \cite{fgb93,fgb94}
to the $K^-$ atomic data using a purely
phenomenological nonlinear density dependent (DD) optical potential
yield a rather different value for its real part in the nuclear interior,
 Re~$V_{opt}=-$200$\pm$20~MeV.
The effect of the $\Lambda$(1405)
is taken into account in these fits implicitly by imposing
 the low density theorem.

In Sec.~\ref{sec:DD} we briefly review the phenomenological fit to
the $K^-$ atomic data using a DD optical potential \cite{fgb93,fgb94}.
The results of applying the chirally motivated coupled channel approach
due to Weise and collaborators \cite{ksw95,wkw96} are discussed in
Sec.~\ref{sec:wkw}.
We find that this microscopic model, in its present form,  fails to
reproduce satisfactorily the $K^-$ atomic data, in spite of the correct
low density limit which is ensured by its success to reproduce the
$\Lambda$(1405) dominance of the near-threshold $K^-p$ physics.
In Sec.~\ref{sec:RMF}
we describe the application of the RMF approach, in which the choice
of the scalar and vector couplings is motivated by SU(3) and by the
quark model. This approach, which violates the low density limit by
disregarding the $\Lambda$(1405), also fails to reproduce satisfactorily
the $K^-$ atomic data. A good fit to the $K^-$ atomic data can be made
only at the expense of a large departure of the fitted RMF scalar and
vector couplings from their underlying theoretical values.
Finally, in Sec.~\ref{sec:RMF+DD} we introduce a hybrid model which
combines the RMF potential form in the nuclear interior
 with a phenomenological DD form at low density. We show that within
 this model, and respecting the
low density limit, it is possible to fit reasonably well the atomic data
with only a moderate departure from the theoretically motivated RMF
couplings. The results of this work are summarized and discussed
in Sec.~\ref{sec:end}.

\section{Density Dependence: Phenomenology}
\label{sec:DD}

The interaction of $K^-$ with the nucleus is described by the Klein-Gordon
(KG) equation of the form presented in Ref. \cite{bfg97}:
\begin{equation}\label{equ:KG1}
\left[ \nabla^2 + k^2 - 2{\varepsilon(k)}(V_{opt} + V_c) + V^2_c\right]
\phi = 0~~ ~~ (\hbar = c = 1)
\end{equation}
where $k$ and $\varepsilon(k)$ are the $K^-$ - nucleus wave number and
reduced energy in the c.m. system, respectively, and $V_c$ is the
Coulomb interaction of the $K^-$ with the nucleus.
The phenomenological DD potential of Friedman {\it et al.}
\cite{fgb93,fgb94} is given at threshold by:

\begin{equation}\label{equ:DD}
2\mu V_{opt}(r) =
 -{4\pi}(1+{\frac{\mu}{m}})b(\rho)
\rho(r) \;\;\; ,
\end{equation}

\begin{equation}\label{equ:b}
b(\rho)= b_0+B_0(\frac{\rho(r)}{\rho_0})^\alpha \;\;\; ,
\end{equation}
\noindent

\noindent
where $\mu$ is the $K^-$-nucleus reduced mass, $b_0$ and $B_0$
are complex parameters determined from fits to
the data, $m$ is the mass of the nucleon and $\rho(r)=\rho_n(r)+\rho_p(r)$
is the nuclear density distribution normalized to the number of nucleons
$A$, and $\rho_0=$0.16 fm$^{-3}$ is a typical central nuclear density.
For $\alpha>0$ and $\rho \rightarrow 0$, the second term on the r.h.s.
of Eq.~(\ref{equ:b}) vanishes and the optical potential of
Eq.~(\ref{equ:DD}) assumes the $t\rho$ form. Furthermore,
for $b_0=-0.15+{\rm i}0.62$~fm  (equal to minus the free $K^-N$
scattering length) the low density limit
is satisfied, so that  the potential is repulsive
in the low density region, reflecting the presence of the subthreshold
$\Lambda$(1405) unstable bound state. Since any fit to the
data yields Re~($B_0 +b_0) > 0$, such potentials become attractive
in the nuclear interior. For the best-fit DD potential, the transition
from repulsion to attraction occurs near
$\rho$/$\rho_0$ = 0.15, the depth of this attractive $V_{opt}$ at $\rho_0$
is then about 190~MeV. For a detailed discussion of the DD potentials
as well as of the $\chi^2$ fits to the kaonic atom data see
Ref.~\cite{bfg97}. Here we note only that the full data base for kaonic
atoms, containing 65 data points over the whole of the periodic table,
was used in the fits, which yielded  $\chi^2$ per point
around 1.5, representing good fits to the data.
Such fits will serve as a reference to the quality
of fits attainable with other models, as is discussed below.
For comparison we note that the best fit (everywhere attractive) $t\rho$
potential is considerably shallower than the best fit DD, but with
$\chi^{2}/N \approx 2.0$ it is also significantly inferior to the latter.

\section{Density Dependence: Theory}
\label{sec:wkw}

\begin{figure}
\epsfig{file=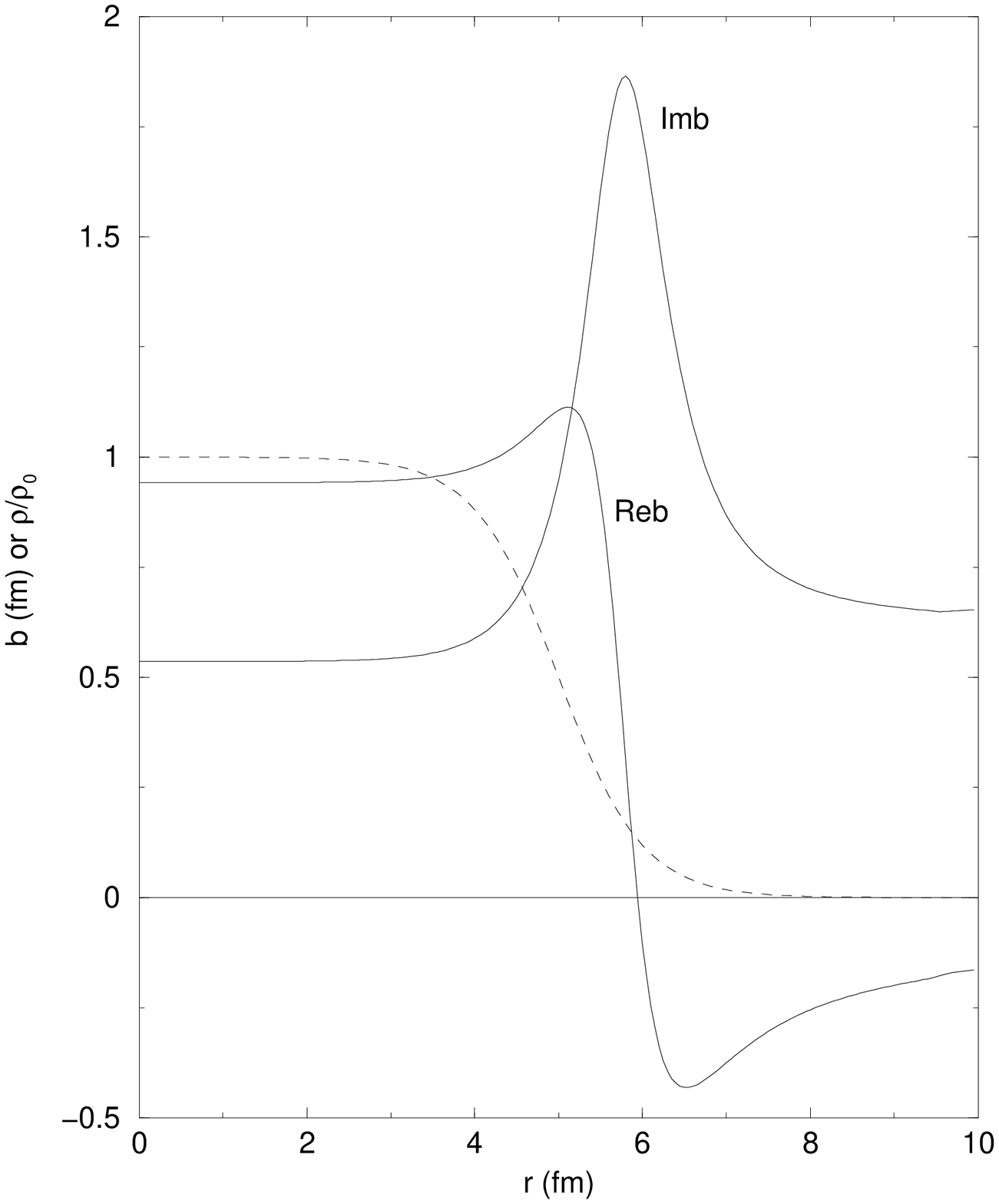,height=140mm,width=120mm,
bbllx=35,bblly=92,bburx=516,bbury=670}
\caption{Solid curves: $b(\rho)$ from the coupled
channel model [11], for a hypothetical nucleus
whose density distribution is shown by the dashed curve.}
\label{fig:Ales}
\end{figure}

Weise and collaborators (for a recent review see Ref.~\cite{ww97}) iterated
the lowest order chiral effective interaction between the pseudoscalar
meson octet and the baryon octet in a Lippmann-Schwinger (LS) coupled
channel matrix equation. For $S=-$1,  the  parameters of
the model  were fitted \cite{ksw95} to
both the low energy $\bar KN$ observables
and to the $\pi\Sigma$ spectrum which provides the major
evidence for the $I$=0 $\Lambda$(1405) $\bar KN$ unstable bound state
27 MeV below the $K^-p$ threshold.
The LS coupled channel equations were then solved in the
nuclear medium \cite{wkw96}, taking into account the Pauli
exclusion principle in
the intermediate nucleon states, as a function of the Fermi momentum
$p_F$ within the Fermi gas model. Fermi motion and nucleon binding
effects were also considered, but turned out to play only a secondary
role. We have reproduced in the present work this calculation
 and  Fig. \ref{fig:Ales} shows the in-medium
isospin-averaged $\bar KN$ threshold scattering amplitude $b$, which is
a function of the corresponding local nuclear density $\rho$.
 It is instructive to plot $b(\rho)$
as a function of position for a hypothetical nucleus with  density
represented by a Fermi function with a radius parameter of 5 fm,
a diffuseness parameter of 0.5 fm, and a central density of 0.16 fm$^{-3}$.
The $K^-$ nucleus optical potential, for a self-conjugate ($N=Z$) nucleus,
is then given by Eq.~(\ref{equ:DD}) in terms of $b(\rho)$.

It is clear from Fig.~\ref{fig:Ales} that Re $V_{opt}$ changes from
repulsion (due to the subthreshold $\Lambda$(1405)) in the
limit of $\rho$=0, where it satisfies the low density theorem with
$b(\rho=0)$ given by the free-space $\bar KN$ threshold scattering
amplitude, to attraction beginning near $\rho$/$\rho_0$ = 0.1, where
$\rho_0$ = 0.16 fm$^{-3}$.
The depth of this attractive real potential at $\rho_0$ is then about
120 MeV. These observations remain in place also when refined
variants \cite{kww97,or98} of the model are used.

For Im~$V_{opt}$, which is due to the one-nucleon absorption processes
$K^{-}N \rightarrow \pi Y$, we note that it deviates significantly from
linearity in the density $\rho$, precisely at the low density region where
Re $V_{opt}$ changes sign, owing to the resonance-like form of Im~$b(\rho)$.
This is the density region where the coupled channel generated $\Lambda$(1405)
moves through the $K^-$ threshold and, therefore, it is there where it
exercises a strong influence on the $K^-$ atomic states.

For nuclei with $N>Z$, the generic form $b(\rho)\rho(r)$ in the
potential (\ref{equ:DD}) actually stands for $b_p(\rho_p)\rho_p(r)
+b_n(\rho_n)\rho_n(r)$. This more general potential was used in the
KG equation~(\ref{equ:KG1}) to evaluate $K^-$ atomic shifts and widths
for the same data set employed in Sec.~\ref{sec:DD}.
The value of $\chi^2$ per point is then close to 15, representing very poor
agreement between predictions and experiment. Applying fudge factors
to the density dependent amplitudes we can obtain reasonably good fits
to the data only at the expense of introducing unphysically large
modifications of the elementary amplitudes, e.g. by a factor of 0.1
for the $K^-p$ and by a factor of 2.4 for the $K^-n$ interactions.
The results are essentially unchanged if we use the present microscopic
approach only for the real potential while employing
a phenomenological approach for the imaginary potential, or vice versa.
It is therefore concluded that although the correct low density behavior is
built into the model, consistently with the effects of the $\Lambda$(1405)
resonance, this microscopic model fails to describe the interaction of
$K^-$ with nuclei at threshold.
Comparing the optical potential of this model to the successful
purely phenomenological model of Sec. \ref{sec:DD}, it is apparent
that fitting to the data excludes a sizable enhancement of Im $b(\rho)$
in the outer nuclear surface and, furthermore, the presence of two
distinct extrema of Re $b(\rho)$ in this region.

\section{Relativistic Mean Field}
\label{sec:RMF}

In the derivation of Re $V_{opt}$ within the RMF approach we make use
of the standard Lagrangian for the nucleons with the linear
parametrization (L) of Horowitz and Serot \cite{hs81}.
For comparison, we quote below results also for the non-linear
parametrization (NL1) due to Reinhard {\it et al.}  \cite{rufa}.
The antikaon sector is incorporated
into the model by using the Lagrangian density of the form
\cite{sm96}:
\begin{equation}
{\cal L}_{K} = \partial_{\mu}\overline{\psi}\partial^{\mu}\psi -
m^2_K\overline{\psi}\psi
- g_{\sigma K}m_K\overline{\psi}\psi\sigma
 - ig_{\omega K}(\overline{\psi}\partial_{\mu}\psi {\omega}^{\mu} -
\psi \partial_{\mu}
\overline{\psi}{\omega}^{\mu})
+(g_{\omega K}{\omega}_{\mu}
)^2
\overline{\psi}\psi.
\end{equation}
The Lagrangian ${\cal L}_K$ describes the interaction of the antikaon field
($\bar{\psi}$) with the scalar ($\sigma$) and vector ($\omega$)
isoscalar fields.
The corresponding equation of motion for $K^-$ in a $Z = N$ nucleus can be
expressed by the KG equation (\ref{equ:KG1}) with the real part of the
optical potential given at threshold by:

\begin{equation}\label{equ:VOP1}
{\rm Re }\;V_{opt}={{m_K}\over{\mu}}({1\over{2}}S - V - {{V^2}\over{2m_K}})
\;\;\; ,
\end{equation}
where $S = g_{\sigma K}\sigma(r)$ and $V = g_{\omega K}\omega_0(r)$ in terms
of the mean isoscalar fields.
Similar considerations have been made recently for the interaction of kaons
with nuclei \cite{fgm97}.
Note that for antikaons, the vector potential $V$ contributes attraction,
just opposite to its role for kaons and nucleons. Each of the three
terms on the r.h.s. of Eq. (\ref{equ:VOP1}), thus,
gives rise to attraction. Consequently,
it becomes impossible to satisfy the low density limit which requires
that, due to the subthreshold $\Lambda$(1405), Re~$V_{opt} > 0$ as
$\rho \rightarrow 0$.
For nuclei with $N>Z$, the potential should include also an isovector
part due to the interaction of the $K^-$ with the $\rho$ meson field.
However, this was omitted from the present calculations as
it was found to have marginal effect in previous analyses of kaonic atoms
\cite{fgb93,fgb94,bfg97}.

In order to avoid calculating RMF real potentials for the 24 different nuclei
contained in the full kaonic atoms data base \cite{fgb94},
we have first confirmed
that using a reduced data set of
only 11  points for carefully selected four atoms,
namely O, Si, Ni and Pb, produces all the features observed in the
potentials obtained from fits to the full data set,
in agreement with the results shown in Table 2 of Ref.~\cite{f98}.
As a guide we used
fits of the phenomenological DD potential of Sec.~\ref{sec:DD} to this
reduced data set.
Since the RMF approach does not address the imaginary
part of the potential, the latter has to be treated on a more
empirical basis, as done for example in recent RMF treatments
of hyperon-nucleus interactions at intermediate
energies \cite{cjm94}, and of $\Sigma^-$ atoms \cite{mfgj95}. Therefore,
for RMF real potentials of the form (\ref{equ:VOP1}),
the imaginary part of Eq.~(\ref{equ:DD}) was used within
a phenomenological DD form, and also within the microscopic approach
of Sec.~\ref{sec:wkw}, and its parameters were fitted to the data.

\begin{figure}
\epsfig{file=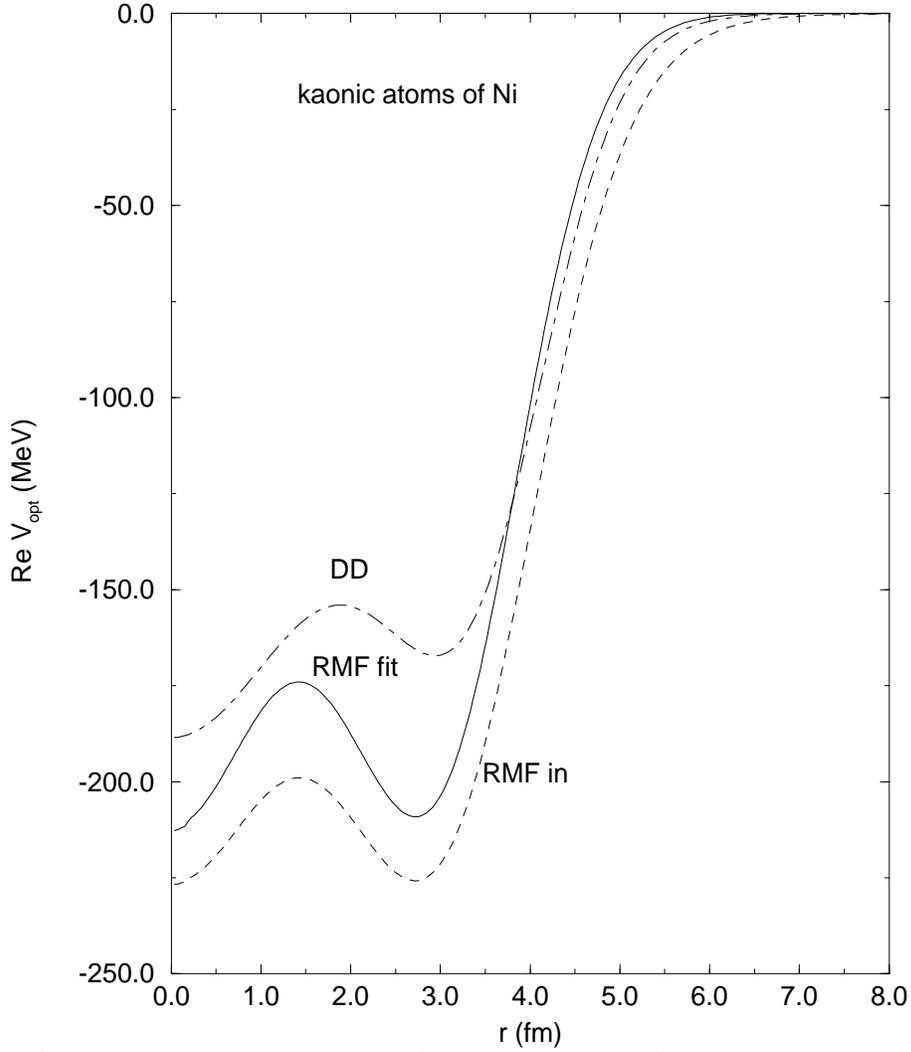,height=140mm,width=120mm,
bbllx=12,bblly=91,bburx=518,bbury=670}
\caption{Real optical potentials for kaonic atoms of Ni: dashed curve for the
RMF(L) input potential (only Im~$V_{opt}$ adjusted), solid curve for the
RMF(L)-adjusted potential and dot-dashed curve for the phenomenological
DD potential.}
\label{fig:Nipotl}
\end{figure}

In order to construct the RMF Re~$V_{opt}$ of Eq.~(\ref{equ:VOP1}) in terms
of the scalar $S$ and the vector $V$ potentials, one needs to specify
$\alpha_{\sigma}$ and $\alpha_{\omega}$, where $\alpha_m = g_{mK}/g_{mN}$.
The quark model (QM) choice $\alpha_{\sigma}=\alpha_{\omega}={1 \over 3}$,
as expected from naively counting the nonstrange quarks in the $\bar K$ with
respect to those in the nucleon \cite{br96}, leads to a very poor fit to
the data, as is clearly seen in Table~\ref{tab:res} for this RMF-in potential
(other parametrizations of Im~$V_{opt}$ led to invariably poor fits).
Therefore, in the next stage, we treated both $\alpha_{\sigma}$ and
$\alpha_{\omega}$ as free parameters and searched for their best fit values.
In this particular case, the effects of the $\Lambda$(1405) which are
expected at low densities, are not included {\it a priori}.
The best fit potential (denoted by RMF-fit in Table~\ref{tab:res})
consists of a strongly attractive vector potential and a strongly
{\it repulsive} scalar potential (note the minus sign for $\alpha_{\sigma}$).
The optical potential Re~$V_{opt}$ corresponding to these coupling
ratios is attractive in the nuclear interior with the depth of
about 190~MeV, in close agreement with the DD potentials \cite{bfg97}.
The solid curve in Fig.~\ref{fig:Nipotl} shows this best fit
RMF(L) potential for Ni.
Also shown in this figure are the DD potential for this example of Ni and the
QM input RMF(L) potential with $\alpha_{\omega}=\alpha_{\sigma}={1 \over 3}$.
The best fit RMF potential is steeper in the surface region than the input
RMF potential, and it becomes, in fact, repulsive at large radii.
Kaonic atoms data thus force the
RMF potential to {\it qualitatively} satisfy the low density limit of
being repulsive at large radii. The sizable departure of the fitted
values of the parameters $\alpha_{\sigma}$ and $\alpha_{\omega}$
from those of the input QM values,
as well as the sharp decrease of $\chi^2/N$ associated with the RMF fit
(see Table \ref{tab:res}),
reflect {\it a posteriori} effects of the $\Lambda$(1405) at low densities.
Various choices for the functional form of Im~$V_{opt}$ were tried,
leading to qualitatively similar results regarding the reversal of $S(r)$
from attraction to repulsion. The imaginary part of the potential used
in the RMF fits specified in Table \ref{tab:res} and shown in
Fig.~\ref{fig:Nipotl} was taken in the $t\rho$ form.
It is also seen from Table~\ref{tab:res}
that Im $V_{opt}$ is well determined for the RMF-fit potential,
with a depth  of about 60 MeV, close
to the value expected from the low density limit. This feature
remained essentially unchanged in the subsequent fits discussed below.

\section{Hybrid Model}
\label{sec:RMF+DD}

The existence of the $\Lambda$(1405) resonance poses a difficulty
for the RMF approach if the parameters $\alpha_m$ are allowed to deviate
only moderately from the values suggested by
the underlying hadron symmetries.
It is also clear that a mean field model cannot be applied reliably
in the low density region where the $K^-$ interaction with the nuclear
medium is affected by the $\Lambda$(1405). In this region, since no
theoretical model has been shown to be successful,
 we  adopted
the  DD  potential of Friedman {\it et al.}
\cite{fgb93,fgb94}, as described in Sec.~\ref{sec:DD}, which takes the
effect of the $\Lambda$(1405) hyperon
into account at least phenomenologically by imposing the low density limit.
On the other hand the RMF description is well justified within the nuclear
interior, for densities larger than about 0.2$\rho_0$, where the effect of
$\Lambda$(1405) can be neglected as demonstrated in Refs.~\cite{wkw96,k94}.
 We have therefore combined the two approaches into
a hybrid model as follows.

In the hybrid model, the functional RMF form Eq.~(\ref{equ:VOP1}) is used
in the nuclear interior for Re~$V_{opt}$, whereas the purely phenomenological
DD form Eq.~(\ref{equ:DD}) is used in the surface of the
nucleus and beyond. The radius $R_M$ where the two forms are
matched to each other is chosen as $R_M\approx R_{1/2} +0.1$~fm, where
$\rho(R_{1/2})=\rho_0/2$.
The sensitivity to the choice of $R_M$ was checked and found to be small.
The value chosen corresponds to the
minimal spread of the fitted coupling constant ratio
$\alpha_{\sigma}$ for the various nuclei. The density
$\rho(R_M)$ is sufficiently high to justify using the RMF approach,
and sufficiently low so that the atomic data are still sensitive
to the RMF form. Figure~4 of Ref.~\cite{bfg97}, and in particular
Fig.~3 of Ref.~\cite{f98} for Ni, show that by analyzing
kaonic atoms one determines the real part of the $K^-$ nucleus DD
optical potential up to $\rho=0.9\rho_0$.
This is well above the density at which the RMF form takes
over in the present approach.

Fitting to the $K^-$ atomic data, the RMF vector coupling constant ratio
$\alpha_{\omega}$ was kept fixed, guided by theoretical considerations,
and the RMF scalar coupling constant ratio $\alpha_{\sigma}$
was varied together with the
parameters of the DD real potential form (for $r>R_M$).
For the imaginary potential, since the RMF approach does not provide
any specific prescription, the purely phenomenological DD potential
form was used throughout.
For the coupling constant $g_{\omega K}$ we used either the constituent
QM value $\alpha_{\omega}={1 \over 3}$, or
the SU(3) relation $2g_{\omega K}=g_{\rho \pi}=6.04$ (denoted by
SU(3) in Table~\ref{tab:res}). Table~\ref{tab:res} shows
that the various hybrid RMF+DD potentials describe the data reasonably well
with $\chi^2/N$ values of $1.4 - 1.5$.

\begin{figure}
\epsfig{file=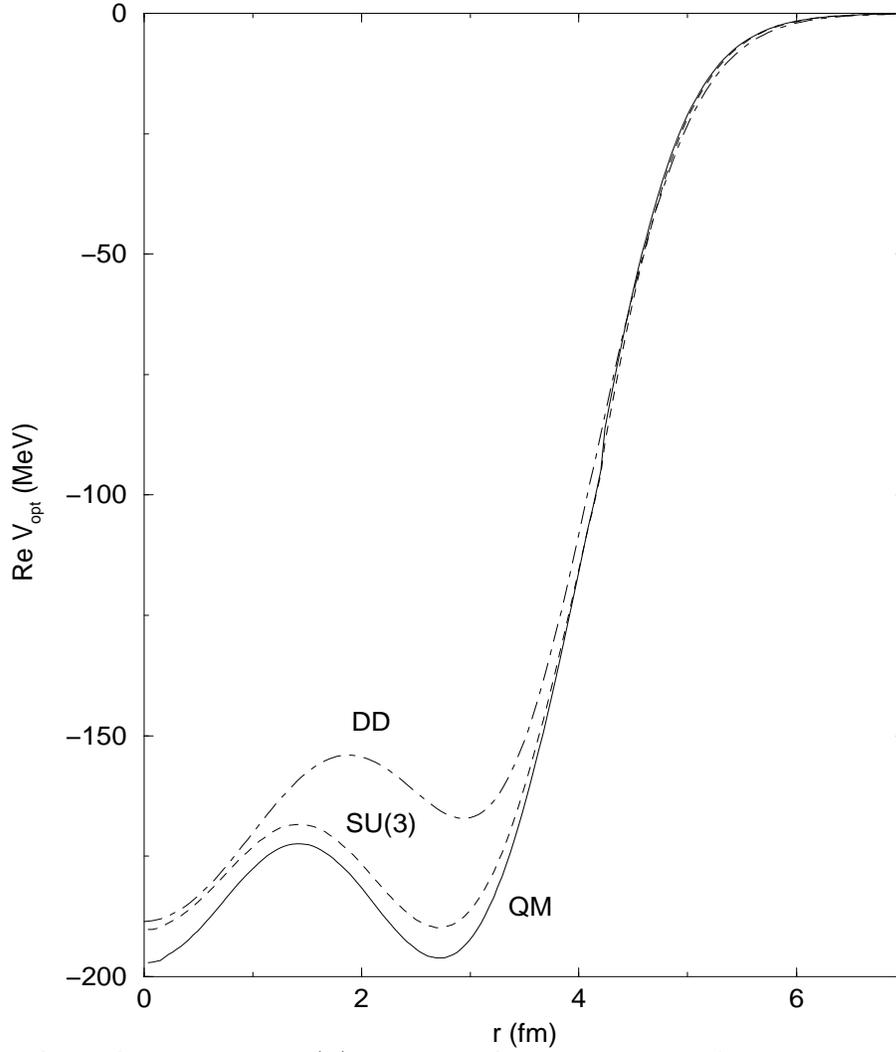,height=140mm,width=120mm,
bbllx=31,bblly=92,bburx=509,bbury=670}
\caption{Combined RMF(L)+DD best fit real potentials for Ni: solid
curve for the QM version, dashed curve for the SU(3) version, see text.
Also shown (dot-dashed) is the phenomenological DD potential.}
\label{fig:NiPb}
\end{figure}

Figure \ref{fig:NiPb} shows the hybrid RMF(L)+DD best fit real potentials
for Ni, for the QM and SU(3) options. The DD potential is also
shown, for comparison. It is seen that
the depths of the hybrid real potentials in the nuclear interior
for the different values of $\alpha_{\omega}$ are about 185~MeV and
are very close to each other, and also to the depth of the purely
phenomenological DD potential. We note that omitting the $V^2$ term
in Eq.~(\ref{equ:VOP1}), and refitting the $K^-$ atomic data, the resulting
depths are smaller than the RMF(L)+DD depths shown in Fig. \ref{fig:NiPb}
by less than 10 MeV.

Figure~4 shows the dependence of the RMF+DD best
fit real potentials for Pb on the type of RMF model used:
the linear model L~\cite{hs81}, and the nonlinear models
NL1~\cite{rufa} and NLS~\cite{sharma}.  The latter nonlinear model,
due to Sharma {\it et al.}, was shown \cite{slr} to fit particularly well
the Pb isotopes, and it is therefore gratifying that it yields a very
similar $V_{opt}$ to that for the linear model.
The QM value $\alpha_{\omega}=1/3$ was used throughout these best fit
potentials.

It is interesting to note the following expression for the $\bar K$
real potential (attractive) depth,

\begin{equation}\label{equ:KDEPTH}
U^{(\bar K)} = {1\over{2}}\alpha_{\sigma}U_{S}^{(N)} +
\alpha_{\omega}U_{V}^{(N)} + {(\alpha_{\omega}U_{V}^{(N)})^{2}\over{2m_K}}
\;\;\; ,
\end{equation}
in terms of the nucleon vector and scalar potential depths, respectively,
which for the linear model L in the mean field approximation are given by

\begin{equation}\label{equ:NDEPTH}
U_{V}^{(N)} = \rho_{V}{{g_{\omega N}^2}\over{m_{\omega}^2}} \; , \;\;\;
U_{S}^{(N)} = \rho_{S}{{g_{\sigma N}^2}\over{m_{\sigma}^2}} \;\;\; ,
\end{equation}
where $\rho_V$ and $\rho_S$ are the nuclear vector and scalar densities,
respectively, evaluated at nuclear matter density.
Analogous expressions for the potential depths of hyperons in nuclear
matter were discussed in Refs.~\cite{gm91,sdggms94}.
For $\rho_{V} = \rho_{0}$ and $\rho_{S} = 0.9 \rho_{0}$,
with $\rho_{0} = 0.16$ fm$^{-3}$, Eq.~(\ref{equ:KDEPTH}) yields for model L:

\begin{equation}\label{equ:185MeV}
U^{(\bar K)} = {\rm 189}\;{\rm MeV}\;\;{\rm (QM)}\;, \;\;\;
               {\rm 181}\;{\rm MeV}\;\;{\rm (SU(3))}\;\; ,
\end{equation}
in excellent agreement with the results of the fits shown in
Figs.~\ref{fig:NiPb} and 4. Thus, for a given model for $\alpha_{\omega}$,
fitting to the $K^-$ atomic data yields a value for $\alpha_{\sigma}$
which results in $U^{(\bar K)} \approx 185$~MeV. This depth $U^{(\bar K)}$
clearly has more physical content than the separate values $\alpha_{m}$,
given in Table~I for the coupling-constant ratios.
We note in passing that this approximate model independence of
$U^{(\bar K)}$  is not shared by the ($S=$1) K-nucleus potential depth
$U^{(K)}$
which can be derived from Eq.~(6) reversing the sign of $\alpha_{\omega}$.
The QM choice yields 66~MeV {\it repulsive} depth, whereas the SU(3)
choice yields 14~MeV {\it attractive} depth. Allowing $\alpha_{\omega}$
to depart moderately, say, from the QM value, it is possible to get
a comparable $K^-$ atomic fit which yields a repulsive depth $U^{(K)}$
at threshold of about 30~MeV, as expected from the scattering length
approximation (see Eq.~(11) of Ref.~[2]). The value of $U^{(\bar K)}$
for this fit remains remarkably close to the value given by Eq.~(8).

The present values of $\alpha_{\omega}$ and $\alpha_{\sigma}$ are not
directly related to the ones encountered in calculations of two-body $KN$
phase shifts. For example, the OBE potential model A of B\"{u}ttgen {\it et al.}
\cite{bhmsw90}, which uses the Bonn potential coupling constants $g_{mN}$
in the nucleonic sector, treats $\alpha_{\omega}$ and $\alpha_{\sigma}$
as free parameters. Although their value of $\alpha_{\omega} = 0.312$ is
very close to the QM value ${1 \over 3}$ adopted here in one of the
hybrid model fits,
the magnitude of $g_{\omega N}$ is somewhat larger in their calculation
than that used in the RMF calculation of Horowitz and Serot \cite{hs81}
adopted in the present work.
Hence the vector contribution to $U^{(\bar K)}$
would be correspondingly larger than the vector contribution in our
calculations. However, their value of $\alpha_{\sigma} = 0.158$ is
considerably smaller than the values obtained in our hybrid model fits,
so that the scalar contribution to $U^{(\bar K)}$ is almost
negligible, less than 20 MeV, for their parameters. Altogether, the
parameters of model A of Ref.~\cite{bhmsw90} would yield
$U^{(\bar K)} = 175$~MeV, fortuitously close to our hybrid model fit
values of Eq.~(\ref{equ:185MeV}).
Subsequent studies by the J\"{u}lich group of the ${\bar K}N$
system \cite{mhs90} and of the $KN$ system \cite{hdhps95} relied on model B
of Ref.~\cite{bhmsw90}, using an additional scalar meson exchange with
unspecified G-parity nature. It is beyond the scope of the present paper
to discuss these works.

\section{Conclusion}
\label{sec:end}

To summarize, by testing several models of the  $K^-$ nucleus
interaction, we have found that only the hybrid model RMF+DD optical
potentials lead to a good agreement with the existing atomic data. These
potentials have an attractive real part in the nuclear interior
with a depth of 180$\pm$20~MeV. This result confirms previous
phenomenological DD analyses \cite{fgb93,fgb94,bfg97} where the density
dependent potentials are constrained to respect the low density limit,
thus including implicitly effects of the $\Lambda$(1405) resonance.

The failure of the chirally motivated coupled channel model \cite{ksw95}
to produce reasonable agreement with the $K^-$ atomic data is disturbing,
since the input free space $t(\rho=0)$ to the $K^-$ nucleus potential
$V_{opt} = t(\rho)\rho$ is constrained in this model by the available
low energy $\bar KN$ experimental data. Recall that the model generates
dynamically the subthreshold $\Lambda$(1405) unstable bound state which
is believed to dominate the low energy $K^-p$ physics. The parameters
used in Ref.~\cite{wkw96} have been updated in Ref.~\cite{kww97};
furthermore, the $\bar KN - \pi Y$ coupled channel model \cite{ksw95}
has been recently extended to include also $\eta Y$ and $K\Xi$ channels
\cite{or98}. Yet, the depth of the resulting $K^-$ nucleus optical
potential has remained low, about 100 MeV according to Ref.~\cite{ro99}.
However, the spectacular failure of the chirally motivated coupled
channel models ($\chi^2/N \approx 15$) cannot be linked just to the depth
of the potential in the nuclear interior, since the $t\rho$ potential
which fits relatively well  ($\chi^2/N \approx 2$) the $K^-$ atomic data
is also found to have a depth of about 100~MeV~[12].
In summary, it is fair to state that these model extrapolations
of the ${\bar K}N$ dynamics at threshold, from free space to finite density,
fail to pass a quantitative test of $K^-$ atoms.

The optical potential $t(\rho)\rho$ is the first order term in a multiple
scattering expansion of $V_{opt}$ (e.g. Ref.~\cite{ek80}). Higher order
terms are not necessarily negligible. The second order term is due to
nuclear pair correlations and has been studied recently by two groups.
Waas {\it et al.} \cite{wrw97} evaluated the effect of Pauli and of short
range correlations and found it to be of a relatively medium sized
{\it repulsive} nature. Pandharipande {\it et al.} \cite{ppt95},
in a calculation geared to high density neutron stars,
found larger repulsive contributions. The leading power of
the density with which nuclear correlations affect $V_{opt}$
is $\rho^{4/3}$.
We note, however,
that the purely phenomenological best fit DD Re~$V_{opt}$ specified in
Table \ref{tab:res}
is dominated in the nuclear interior
by an {\it attractive} $\rho^{4/3}$ term.
Thus, the above two models \cite{wrw97,ppt95} give rise to a considerably
shallower $K^-$ nucleus potential in the nuclear interior than the
best fit DD potential.
This discrepancy poses a major
theoretical challenge. One cannot rule out that the multiple
scattering conventional approach to $V_{opt}$ breaks down for the $K^-$
nucleus system, unless a self consistency requirement similar to that
considered very recently by Lutz \cite{l98} is imposed on the $K^-$
optical potential. However, this self-consistent treatment gives even shallower
$K^-$ nucleus potential than due to the coupled channel model of
Refs.~\cite{ksw95,wkw96}.
Finally, we mention attempts to construct
microscopically the density dependence of $V_{opt}$, considering
explicitly $\Lambda$(1405) degrees of freedom, by using a $\Lambda$(1405)
particle - nucleon hole model for the nuclear excitations involved
(Ref.~\cite{mht94} and references cited therein). Limited sets of $K^-$
atomic data have been fitted in such models, at the expense of
introducing new parameters that have to do with the hypothesized complex
mean field experienced by the $\Lambda$(1405). The simplest term that
must be added to the underlying Lagrangian, in order to enforce such
attempts, is an effective four Fermi interaction involving both $N$ and
$\Lambda$(1405), with unknown coupling constant. The problematics of
this approach are discussed in Sec. 5 of Ref.~\cite{wrw97}.

The failure of the RMF model, with couplings not too dissimilar to those
motivated by the QM or SU(3), is less of a theoretical concern, since the
resulting $V_{opt}$ which is attractive everywhere ignores the effect of
the $\Lambda$(1405) by violating the low density limit.
It appears that a necessary condition for $V_{opt}$ to produce a good
agreement with the $K^-$ atomic data is that it behaves properly
at low density. This is the prime motivation for introducing the hybrid
RMF+DD model in Sec.~\ref{sec:RMF+DD}.
Previous applications of the RMF approach to the $K^-$
sector (\cite{smb97} and references cited therein plus, quite recently,
the related quark meson coupling model calculation \cite{tstw98})
have not been meaningfully constrained by low energy $\bar KN$ data.
It is worth recalling in this context that the RMF nucleon couplings, also, do
not follow directly from $NN$ data, although they roughly follow from the
Lorentz structure of several OBE $NN$ potential models.
However, for the RMF approach to assume a broad validity in hadronic physics,
one would like the couplings in the $S=-$1 strange sector to be related to
those in the $S$=0 nonstrange sector, as suggested by the underlying symmetries
governing hadron structure, such as the QM and SU(3) used here. This harmonious
viewpoint is only partially successful in RMF studies of hypernuclei \cite{mj94}
and of $\Sigma^-$ atoms \cite{mfgj95}. Similarly, for the coupling ratios
determined by fitting the hybrid RMF+DD model to the $K^-$ atomic data
(Table~\ref{tab:res}), moderate departures occur from a pure adherence
to the QM, or to the SU(3) coupling ratios. We conclude that the hybrid model
potentials represent the most theoretically inclined $K^-$ nucleus
potentials which are worth using for extrapolating into high density matter.

\vspace{8mm}
This research was partially supported by the Israel Science Foundation
(E.F.), by the U.S.-Israel Binational Science Foundation (A.G.) and by
the Grant Agency of the Czech Republic (A.C. and J.M., grant No. 2020442).
J.M. acknowledges the hospitality of the Hebrew University.
A.C. thanks T. Waas for useful information on the subject matter of
Ref.~\cite{wkw96}.

\begin{figure}
\epsfig{file=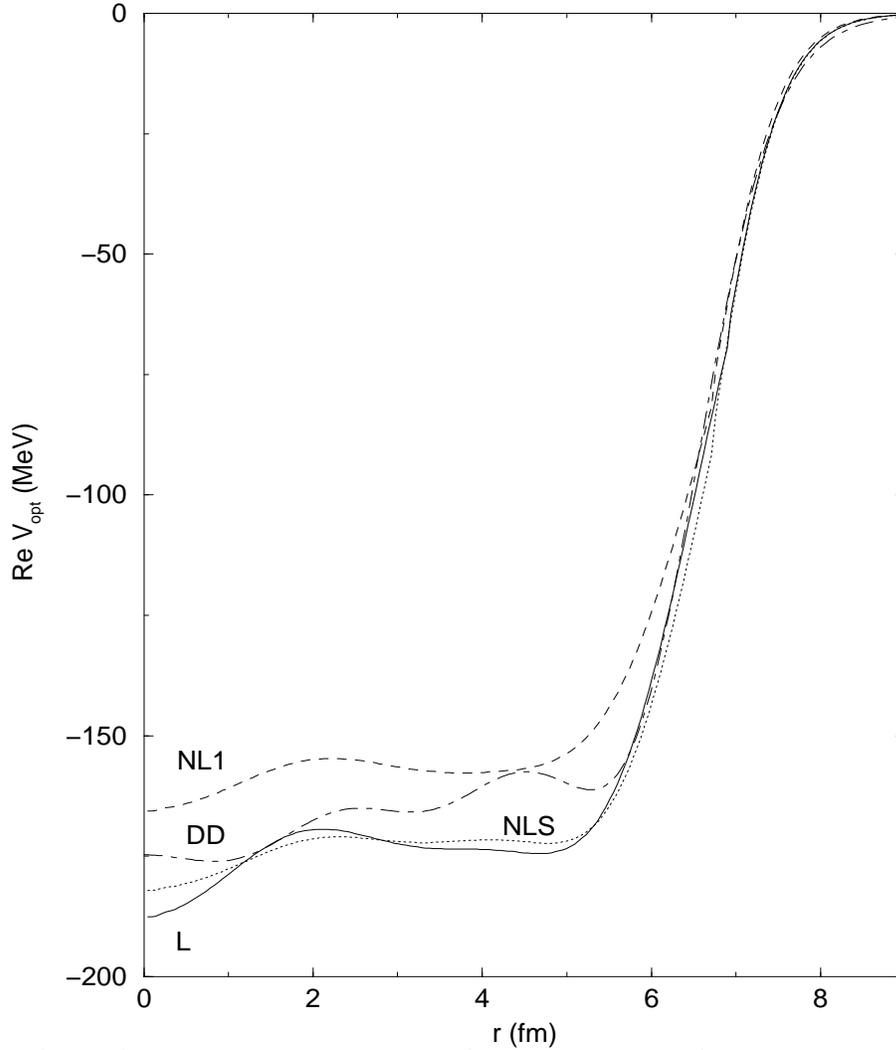,height=140mm,width=120mm,
bbllx=31,bblly=92,bburx=509,bbury=670}
\caption{Combined RMF+DD best fit real potentials for Pb using the
QM version, see text. The solid curve is for model L [15], the dashed
curve is for model NL1 [16] and the dotted curve is for model
NLS [23]. Also shown (dot-dashed) is the phenomenological DD
potential. }
\end{figure}

\begin{table}
\caption{Parameters of the various optical potentials
that fit the $K^-$ atom data (see text). $b_0$ and $B_0$
are in fm, underlined quantities were held fixed during the fits,
$\alpha$=1/3 and 0.31$\pm$0.08 for the hybrid and DD models, respectively.}
\label{tab:res}
\begin{tabular}{lcccccc}
 & $\alpha_{\omega}$ & $\alpha_{\sigma}$ &$b_0$&$ReB_0$&$Im B_0$ &
$\chi^2$/N \\ \hline
RMF(L)-in & $\underline{1\over 3}$& $\underline{1\over 3}$& i(1.12$\pm$0.50)& --- &
 --- & 18.0 \\
RMF(L)-fit & 0.75$\pm$0.03&
$-$0.73$\pm$0.09& i(0.48$\pm$0.09)& --- & --- & 1.53 \\
RMF(L)+DD & $\underline{{\rm QM}}$& ${1\over 3}$(0.60$\pm$0.01)&
 \underline{$-$0.15+i0.62} &
1.95$\pm$0.06&$-$0.19$\pm$0.04 & 1.40 \\
RMF(NL1)+DD & $\underline{{\rm QM}}$& ${1\over 3}$(0.70$\pm$0.03)&
 \underline{$-$0.15+i0.62} &
1.78$\pm$0.05&$-$0.20$\pm$0.05 & 1.39 \\
RMF(L)+DD & $\underline{{\rm SU(3)}}$ & ${1\over 3}$(1.21$\pm$0.04)&
\underline{$-$0.15+i0.62} &
2.01$\pm$0.06&$-$0.19$\pm$0.05 & 1.49 \\
RMF(NL1)+DD & $\underline{{\rm SU(3)}}$ & ${1\over 3}$(1.30$\pm$0.02)&
\underline{$-$0.15+i0.62} &
1.83$\pm$0.05&$-$0.19$\pm$0.05 & 1.47 \\
DD & --- & --- &\underline{ $-$0.15+i0.62} &
1.79$\pm$0.08&$-$0.22$\pm$0.08 & 1.28 \\
\end{tabular}
\end{table}

\end{document}